\documentclass[aps,prd,twocolumn,notitlepage,nofootinbib,,nobibnotes,superscriptaddress,amsmath,amssymb,preprintnumbers]{revtex4-1}

 \usepackage{url}
\usepackage{graphicx}
\usepackage{dcolumn}
\usepackage{bm}        
\usepackage{amssymb}   
\usepackage{adjustbox}
\usepackage{amsmath,array}
\usepackage{comment}
\usepackage{natbib}
\usepackage{MnSymbol}
\usepackage[hidelinks]{hyperref}
\usepackage{graphicx}
\usepackage{subcaption}
\usepackage{color}
\usepackage     [utf8]                  {inputenc}
\usepackage     [T1]                    {fontenc}
\usepackage     [english]               {babel}
\usepackage{epigraph}
\usepackage{etoolbox}
\usepackage{ dsfont }
\usepackage{physics}
\usepackage{xcolor}
\usepackage{siunitx}
\makeatletter
\patchcmd{\epigraph}{\@epitext{#1}}{\itshape\@epitext{#1}}{}{}
\newcommand*\eqsize{%
\@setfontsize\mysize{9.0}{9.0}%
    }
    \usepackage{multirow}
\makeatother

\usepackage{xfrac}
\usepackage{amsmath,amssymb}
\usepackage{esdiff} 
\usepackage{commath} 
\usepackage{bbm} 
\usepackage{braket}
\usepackage{slashed}

\newcommand{\rmd}{\mathrm{d}}

\newcommand{\rmB}{\mathrm{B}}
\newcommand{\rmN}{\mathrm{N}}
\newcommand{\rmP}{\mathrm{p}}
\newcommand{\rmh}{\mathrm{h}}

\newcommand{\pbar}{\bar{\rmP}}

\newcommand{\ppbar}{\rmP\pbar}
\newcommand{\BBbar}{\rmB\bar{\rmB}}
\newcommand{\NNbar}{\rmN\bar{\rmN}}

\newcommand{\snn}{\sqrt{s_{NN}}}

\definecolor{oscar}{RGB}{22, 156, 172}

\newcommand{\mref}[1]{\textbf{\textsc{\textcolor{red}{[REF]}}}}

\begin{document}

\date{\today}

\title{The role of proton-antiproton regeneration in the late stages of heavy-ion collisions}

\author{Oscar Garcia-Montero}
\affiliation{Institut f\"{u}r Theoretische Physik,
                     Goethe Universit\"{a}t, Max-von-Laue-Strasse 1,
                     60438 Frankfurt am Main, Germany}
\author{Jan Staudenmaier}
\affiliation{Institut f\"{u}r Theoretische Physik,
                     Goethe Universit\"{a}t, Max-von-Laue-Strasse 1,
                     60438 Frankfurt am Main, Germany}
\affiliation{Frankfurt Institute for Advanced Studies,
				Ruth-Moufang-Strasse 1, 60438 Frankfurt am Main, Germany}
\author{Anna Schäfer}
\affiliation{Institut f\"{u}r Theoretische Physik,
                     Goethe Universit\"{a}t, Max-von-Laue-Strasse 1,
                     60438 Frankfurt am Main, Germany}
\affiliation{Frankfurt Institute for Advanced Studies,
				Ruth-Moufang-Strasse 1, 60438 Frankfurt am Main, Germany}
\affiliation{GSI Helmholtzzentrum f\"{u}r Schwerionenforschung,
                     Planckstr. 1, 64291 Darmstadt, Germany}
\author{Juan M. Torres-Rincon}
\affiliation{Institut f\"{u}r Theoretische Physik,
                     Goethe Universit\"{a}t, Max-von-Laue-Strasse 1,
                     60438 Frankfurt am Main, Germany}
\author{Hannah Elfner}
\affiliation{GSI Helmholtzzentrum f\"{u}r Schwerionenforschung,
             Planckstr. 1, 64291 Darmstadt, Germany}
\affiliation{Institut f\"{u}r Theoretische Physik,
                     Goethe Universit\"{a}t, Max-von-Laue-Strasse 1,
                     60438 Frankfurt am Main, Germany}
\affiliation{Frankfurt Institute for Advanced Studies,
				Ruth-Moufang-Strasse 1, 60438 Frankfurt am Main, Germany}
\affiliation{Helmholtz Research Academy Hesse for FAIR (HFHF), GSI Helmholtz Center, Campus Frankfurt, Max-von-Laue-Straße 12, 60438 Frankfurt am Main, Germany}

\begin{abstract}
We investigate the long-standing question of the effect of proton-antiproton annihilation on the (anti-)proton yield, while respecting detailed balance for the 5-body back-reaction for the first time in a full microscopic description of the late stages of heavy-ion collisions. This is achieved by employing a stochastic collision criterion in a hadronic transport approach (SMASH), which allows to treat arbitrary multi-particle reactions. It is used to account for the regeneration of (anti-)protons via $5\pi\rightarrow \ppbar$. Our results show that a back-reaction happens for a fraction of 15-20\% of all annihilations. Within a viscous hybrid approach Au+Au/Pb+Pb collisions from $\sqrt{s_{NN}}=17.3$ GeV$-5.02$ TeV are investigated and the quoted fraction is independent of the beam energy or centrality of the collision. Taking the back-reaction into account results in regeneration of  half of the (anti-)proton yield that is lost due to annihilations at midrapidity. We also find that, concerning the multiplicities, treating the back-reaction as a chain of 2-body reactions is equivalent to a single 5-to-2 reaction.
\end{abstract}

\maketitle

\section{Introduction}

During the last stage of heavy-ion collisions, hadrons are formed and interact with each other before flying to the detector. In this stage, dynamical non-equilibrium effects are essential to fully describe the observables which characterize the measured hadrons. A good example
is the mismatch of the predicted (anti-)proton yields in thermal models, later alleviated by the inclusion of $\pi$-nucleon interaction terms~\cite{Andronic:2012dm,*Andronic:2017pug,*Stachel:2013zma, Andronic:2018qqt}.

Another mechanism proposed to be the cause of the mismatch is  the dynamical baryonic annihilation~\cite{Becattini:2012sq,*Steinheimer:2012rd}. This was further confirmed by hydrokinetic studies  where agreement with LHC data improved, if baryon-antibaryon ($\BBbar$) annihilation was included in the afterburner~\cite{Karpenko:2012yf}. Nevertheless, these studies did not include the backward reaction (regeneration), which sparked debate on whether detailed balance significantly affects the yields. Previous works hint at non-negligible effects, where a significant fraction of the pairs lost to annihilation are regenerated~\cite{Rapp:2000gy,Pan:2012ne}. Because the annihilation/back-reaction interplay is a dynamical one, one needs a microscopic transport approach which takes into account back-reactions of the annihilation channels. Such an implementation is instrumental to resolve this discussion and determine the impact regeneration has on the proton-antiproton ($\ppbar$) yields. Since the proton yields are most sensitive to the effects of the switching temperature~\cite{Ryu:2017qzn}, a sizeable regeneration will affect the ongoing efforts to constraint the QCD transport coefficients~\cite{JETSCAPE:2020mzn,JETSCAPE:2020shq}.

In this work, we report for the first time, results employing 5-body back-reactions for nucleon-antinucleon ($\NNbar$) annihilation in a transport approach, which restore detailed balance. This in particular allows to quantify the regeneration of (anti-)protons in the late non-equilibrium stages of collisions. Two different treatments for the $5 \leftrightarrow 2$ reactions are presented in this work. First, we extend the stochastic collision criterion introduced in Ref.~\cite{Staudenmaier:2021lrg} to $5 \leftrightarrow 2$ reactions (stochastic treatment). Second, the same overall reaction is handled via intermediate resonances as a chain of two-body reactions (resonance treatment). While the former is theoretically cleaner, the latter is computationally less intensive but restricted to two-body reactions (like the usually employed geometric collision criteria~\cite{Bass:1998ca,Weil:2016zrk}) in order to conserve detailed balance. The comparison of both treatments, furthermore, allows to gauge the validity of employing multi-step reaction chains involving resonances with finite lifetimes for multi-particle reactions.

The treatment using the stochastic collision criterion builds on earlier works employing similar methods as in~\cite{Cassing:2001ds,Seifert:2017oyb,Seifert:2018bwl}, where $\ppbar$ annihilation reactions were studied. However, in these prior works only $3 \leftrightarrow 2$ reactions are used, where the $5\pi$ final state is created by resonance decays from reactions like $\ppbar\rightarrow\rho\rho\pi$. The latter is comparable to our above described  treatment using intermediate resonances.
\begin{figure*}[ht]
 \centering
 \includegraphics[width=0.9\textwidth]{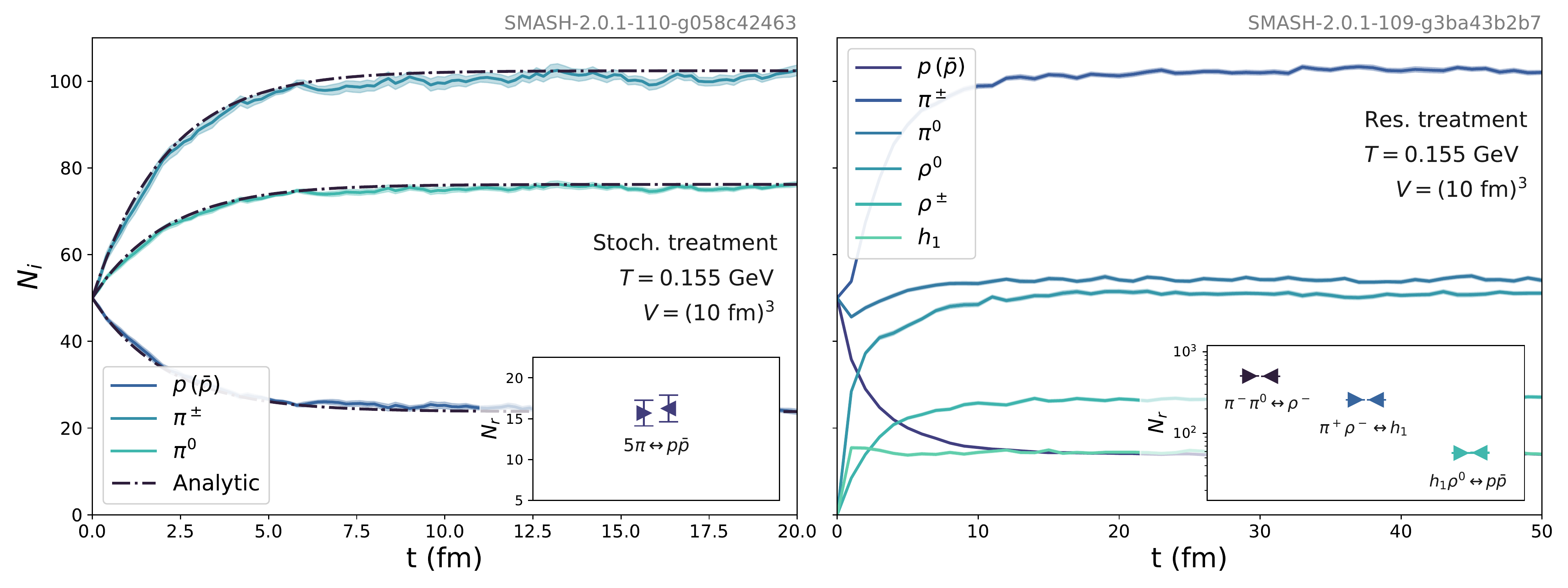}
  \caption{\label{fig:box} Time evolution of particle multiplicities in infinite matter calculations. Left panel: Stochastic treatment applied to $\ppbar \leftrightarrow 5\pi$ reaction. Right panel: Resonance treatment for intermediate $2\leftrightarrow2$ and $1\leftrightarrow 2$ reactions. The analytic result from~\cite{TimMaster} is obtained by solving a deterministic system of rate equations. Inset: Total number of reactions per event ($N_r$) in the forward ($\blacktriangleright$) and backward ($\blacktriangleleft$) directions.}
\end{figure*}
In our setup we find an excellent agreement between the resonance and the stochastic treatment, as well as a stable ratio of regeneration/annihilation reactions across different energies and centralities. Further, we find an increase in the nucleon yields and regeneration probabilities of up to 50$\%$ at midrapidity for all collision energies.

This letter is organized as follows. In Sec.~\ref{sec:model} we briefly give an account of the space-time modeling using a hybrid approach. Then, both reaction treatments are explained Sec.~\ref{sec:ppbar}. In Sec.~\ref{sec:results} we present our results and discussion, and finally we give a brief summary and outlook on Sec.~\ref{sec:conclusions}.

\section{Modeling the evolution}
\label{sec:model}

To extract the behavior of proton-antiproton annihilation and restoration in high energy collisions, we use a hybrid approach to simulate Au-Au collisions at $\sqrt{s_{NN}}=39$ GeV, $\sqrt{s_{NN}}=200$ GeV, as well as Pb-Pb collisions at $\sqrt{s_{NN}}=17.3$ GeV, $\sqrt{s_{NN}}=2.76$ TeV and $\sqrt{s_{NN}}=5.02$ TeV. We employ the SMASH-vHLLE-Hybrid approach \cite{SVH}. 
The initial conditions are provided by the SMASH hadronic transport approach \cite{Weil:2016zrk, SMASH:DOI}, the evolution of the fireball is performed by the 3+1D viscous hydrodynamics code vHLLE \cite{Karpenko:2013wva}, particlization is achieved with the SMASH-hadron-sampler \cite{Karpenko:2015xea,SHS}, and SMASH is again applied for the hadronic afterburner evolution.

Our objective is not to achieve a precise description of experimental data, but to employ values for parameters of the hybrid approach that are known to reproduce bulk properties at high beam energies well  (compatible with ~\cite{JETSCAPE:2020mzn, *JETSCAPE:2020shq}). To limit computational expenses we restrict ourselves to averaged initial conditions, since event-by-event fluctuations are not crucial for the average yields of produced particles. The initial conditions are obtained by averaging 100 nucleus-nucleus collisions for 0-5\%,20-30\% and 40-50\% centrality classes. The particles are propagated and interact until they reach a hypersurface of constant proper time determined by nuclear overlap and set to $\tau_0=0.5$ fm at and above $\sqrt{s_{NN}}=200$ GeV. Upon crossing this hypersurface, Gaussian smearing is applied across all events to obtain the averaged 3+1D initial conditions for vHLLE. For the hydrodynamics evolution we apply the chiral model equation of state \cite{Steinheimer:2010ib}, and use a shear viscosity of $\eta/s = 0.1$ and bulk viscosity of $\zeta/s=0.05$  for all collision energies. The medium is evolved according to viscous hydrodynamics until the energy density drops below $\epsilon_{\mathrm{crit}}=0.5 $ GeV/fm$^{3}$ \cite{Huovinen:2012is}. Subsequently, the Cooper-Frye formula is evaluated on the $\epsilon_{\mathrm{crit}}$-hypersurface to yield particlization of the fluid elements. We sample 2000 events, using the SMASH-hadron-sampler, which then serve as initial conditions for the non-equilibrium afterburner evolution. The hadrons are further propagated in SMASH and the remaining interactions performed until the medium is too dilute.

\section{Proton-antiproton annihilation and back-reaction}
\label{sec:ppbar}

In SMASH, the collision term of the relativistic Boltzmann equation is realized by binary (in-)elastic scatterings as well as the formation and decay of resonances. Additionally, string excitation and fragmentation are employed for highly energetic collisions. In this work we rely on an extension with stochastic rates recently introduced~\cite{Staudenmaier:2021lrg}, which allows for multi-particle reactions. Even though these reactions are implemented for both nucleon isospin states, we focus in our presentation on the species of experimental interest, $\mathrm{p}$ and $\pbar$. Since the dynamics of baryon resonances does not differ significantly from $\ppbar$, one can think of them as a proxy for all $\mathrm{B}\bar{\mathrm{B}}$ annihilation reactions. The latter (excluding nucleons) are realized via string fragmentation for the presented calculations.

Proton-antiproton annihilations are be performed in SMASH by three different methods. First, $\NNbar$ annihilation is realized via string fragmentation (default)~\cite{Mohs:2019iee}. As in all previous works including hadronic afterburner calculations, the back-reaction is unaccounted for and detailed balance is broken. Nevertheless, detailed balance can be restored when handling the $\ppbar$ annihilation via intermediate resonances as a chain of two-body reactions, that characterizes the second method. As such, the annihilation of the nucleon pair is performed via $\ppbar \rightarrow \rmh_1 \rho$, which subsequently decays to $5\pi$ (via $\ppbar \rightarrow \rmh_1 \rho \rightarrow \rho\pi\pi\pi\rightarrow5\pi$). This multi-step process in turn provides the back-reaction $5\pi\rightarrow \ppbar$, needed for the restoration of $\ppbar$ pairs. Detailed balance is conserved for all intermediate and total reactions. This method was used to extract the shear viscosity of hadronic matter in a periodic box in Refs.~ \cite{demirphd,Rose:2017bjz}.
We note that the $\ppbar \rightarrow \rmh_1 \rho$ process is experimentally not well constrained, but this reaction chain construct is essential to treat a 5-body final state. In addition, the reaction is slowed down when introducing intermediate resonances. The impact of finite lifetimes on the final results is gauged by comparison to the direct stochastic treatment.

The third treatment is an extension of the stochastic criterion in SMASH, which allows to treat $5 \leftrightarrow 2$ reactions in a single step, fulfilling detailed balance. For this criterion, the relevant collision term is used to compute a reaction probability. The scattering is performed using a Metropolis-like algorithm. This method bears the advantage that, unlike the geometric criterion, it can be extended to arbitrary $n\leftrightarrow m$ reactions. The reader is referred to~\cite{Staudenmaier:2021lrg} for a more comprehensive description of the stochastic criterion. The collision probability for a 5-to-2 reaction, assuming that the scattering matrix element only depends on the Mandelstam variable $s$, is given by the expression
\begin{equation}
  P_{5 \rightarrow 2} =  g'_1\,g'_2\left[\prod_{f=1}^{5} \frac{1}{g_f 2E_f} \right] \frac{S_{5}}{S'_{2}} \frac{\Delta t}{(\Delta^3 x)^4} \frac{\lambda(s,m_1^{\prime2},m_2^{\prime2})}{\Phi_5} \frac{\sigma_{2 \rightarrow 5}}{4\pi s}\ ,
\label{eq:p52}
\end{equation}
where primed quantities refer to outgoing particles. $g_i$ are the spin degeneracies ($g_\pi=1,g_{\rmP/\pbar}=2$) and $S_i$ the symmetry factors accounting for indistinguishable states in a collision ($S_{\ppbar}=1$, $S_{\pi^+\pi^+\pi^-\pi^-\pi^0}=2!2!$). $E_f$ are the energies of the incoming particles, $\lambda (a,b,c) = (a-b-c)^2-4bc$ is the K{\"a}ll{\'e}n function, $\Phi_5$ is the integrated 5-body phase space, and the cross section $\sigma_{2 \rightarrow 5}$ for the $\ppbar \to 5\pi$ process is taken as follows. 

We assume that the $\ppbar$ inelastic cross section (difference between the experimental total and elastic cross sections) is saturated by multi-pion reactions. This follows from the subleading contributions of other channels as discussed in Refs.~\cite{Koch:1986ud,Dover:1992vj,Cassing:2001ds}. Second, instead of implementing the whole set of processes $\ppbar \rightarrow m\pi$, with $m=2,3...$ pions~\cite{Dover:1992vj}, we consider an effective approach employing a single scattering process emitting an (energy-dependent) average number of pions. For the typical energies of the $\rmP(\pbar)$ in this work, the corresponding $\sqrt{s}$ in the $\ppbar$ annihilation process is close to the two-nucleon mass threshold. In this energy range the average number of produced pions is $m=5$~\cite{Dover:1992vj}, which is the process we implement in this work with a $\sigma_{2 \rightarrow 5} (s)$ corresponding to the inclusive multi-pion cross section, and, from the first argument, equal to the $\ppbar$ inelastic cross section. In Eq.~(\ref{eq:p52}) $\Delta t$ is the time-step size of the dynamical evolution of the system, and $\Delta^3 x$  is the sub-volume where the $5\rightarrow 2$ collision happens. The collision probability is calculated for all possible 5- and 2-particle combinations, making the calculation computationally expensive. Nevertheless, since this treatment is directly based on the Boltzmann equation, it is the theoretically most rigorous treatment available, while at the same time allowing to treat all possible final states. 

\begin{figure}[t]
	\centering
	\includegraphics[width=0.4\textwidth]{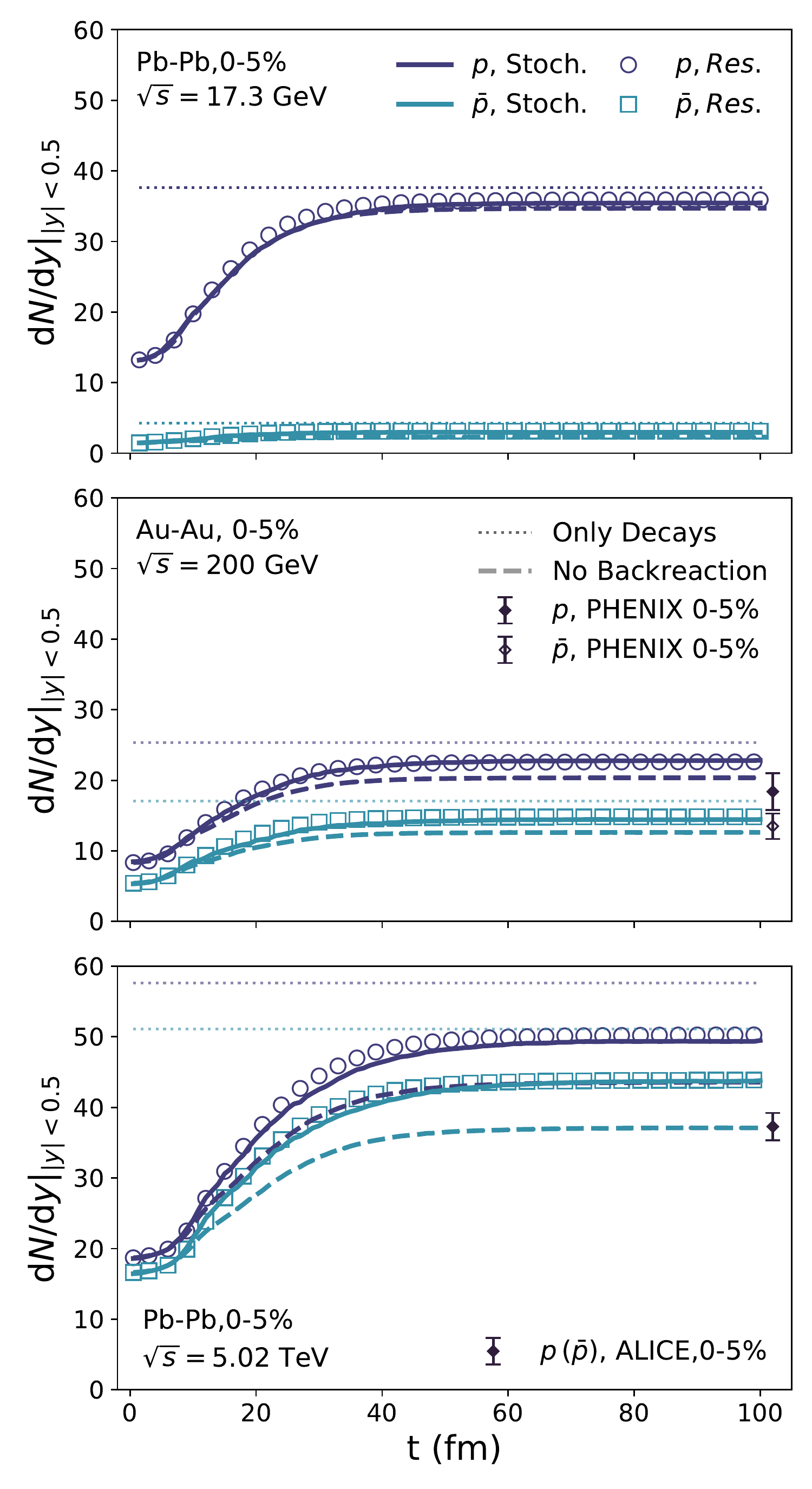}
	\caption{\label{fig:yields}Time evolution of $\rmP$ and $\pbar$ yields at midrapidity for different collision systems and energies: Pb-Pb at $\sqrt{s_{NN}}=17.3$ GeV (top), Au-Au at $\sqrt{s_{NN}}=200$ GeV (middle) and Pb-Pb at $\sqrt{s_{NN}}=5.02$ TeV (bottom). The results from the stochastic approach are depicted by solid lines, while the results from the resonance treatment are given as open symbols. Experimental data from Refs.~\cite{ALICE:2019hno,PHENIX:2003iij} }
\end{figure}
\begin{figure*}[ht]
	\centering
	\includegraphics[width=0.9\textwidth]{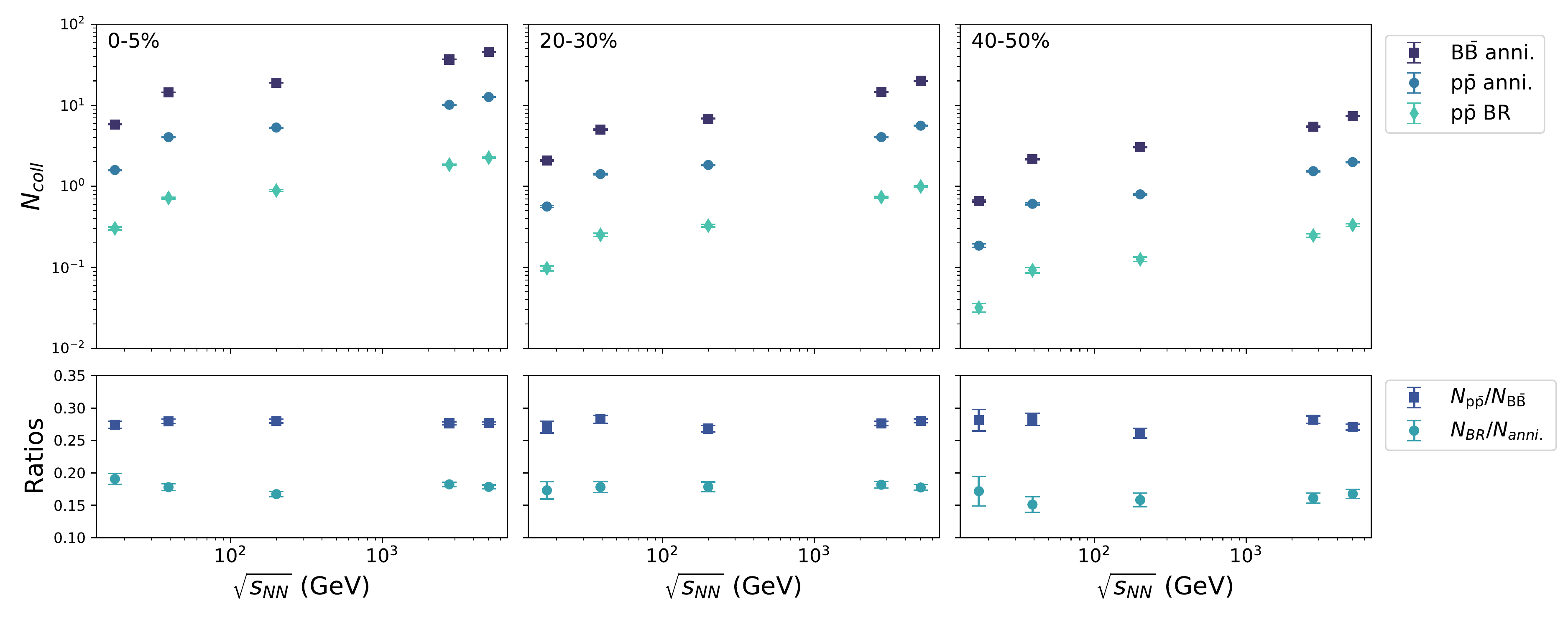}
	\caption{\label{fig:Reactions} (\textit{Above}) Total number of reactions for different systems (Au-Au, $\sqrt{s_{NN}}=39,200$ GeV, Pb-Pb, $\sqrt{s_{NN}}=17.3$ GeV, $2.76$ and $5.02$ TeV.), for different centrality classes (0-5, 20-30 and 40-50$\%$). (\textit{Below}) Ratio of total number of $\ppbar$ to $\BBbar$ annihilation reactions, as well as $\ppbar$ back-reaction (BR) to annihilation reactions.} 
\end{figure*}
We have benchmarked the stochastic treatment by testing equilibration in two infinite matter settings, one employing the stochastic rates for annihilation processes and one employing the intermediate resonances. A $\rm (10~fm)^3$ box with periodic boundary conditions has been initialized in both cases with $50$ nucleons and pions of each isospin state. Fig.~\ref{fig:box} shows the multiplicity evolution for the two cases. While it is observed that all yields equilibrate chemically for both settings, equilibration is significantly faster for the stochastic treatment, with yields not changing after around $7.5$ fm. This is a consequence of the more complex, multi-step treatment involving intermediate resonances with finite lifetimes. The faster equilibration of the medium when employing multi-particle reactions, confirms previous findings for 3-to-2 and 3-to-1 reactions in~\cite{Staudenmaier:2021lrg}. 

The insets of Fig.~\ref{fig:box} demonstrate detailed balance by counting the total number of forward and backward reactions per event. This proves that this calculation is the first one in which detailed balance is achieved for a 5-body reaction in a transport approach. For the stochastic approach of 5-to-2 reactions, the equilibration process is also compared to the solution of analytic rate equations given in Ref.~\cite{TimMaster}, and found to be fully compatible (see Fig.~\ref{fig:box}). Yet, direct comparison to the resonance treatment is not possible as different degrees of freedom are required to realize the full reaction chain.

\section{Results}
\label{sec:results}


In what follows, we track the evolution of the midrapidity \mbox{(anti-)proton} number, $\rmd N/\rmd y$ throughout the afterburner evolution, as well as the number of reactions ($\ppbar$ annihilation and back-reaction) and find an excellent agreement between the stochastic and resonance treatments.
In Fig.~\ref{fig:yields}, we distinguish three scenarios in our calculations: Performing \emph{only decays} after particlization without rescattering (dotted lines), similar to particle production assumed by thermal models~\cite{Andronic:2017pug,Stachel:2013zma}. Secondly, we include rescattering with \emph{no back-reaction} (dashed lines), which shows the maximal effect of $\ppbar$ annihilations (see Refs.~\cite{Steinheimer:2012rd, Becattini:2000jw,Becattini:2014hla}). Finally, the $5\pi\rightarrow \ppbar$ reaction is taken into account with the stochastic and resonance treatments (solid lines and markers, respectively).

\begin{figure}[hb]
   \centering
   \includegraphics[width=0.4\textwidth]{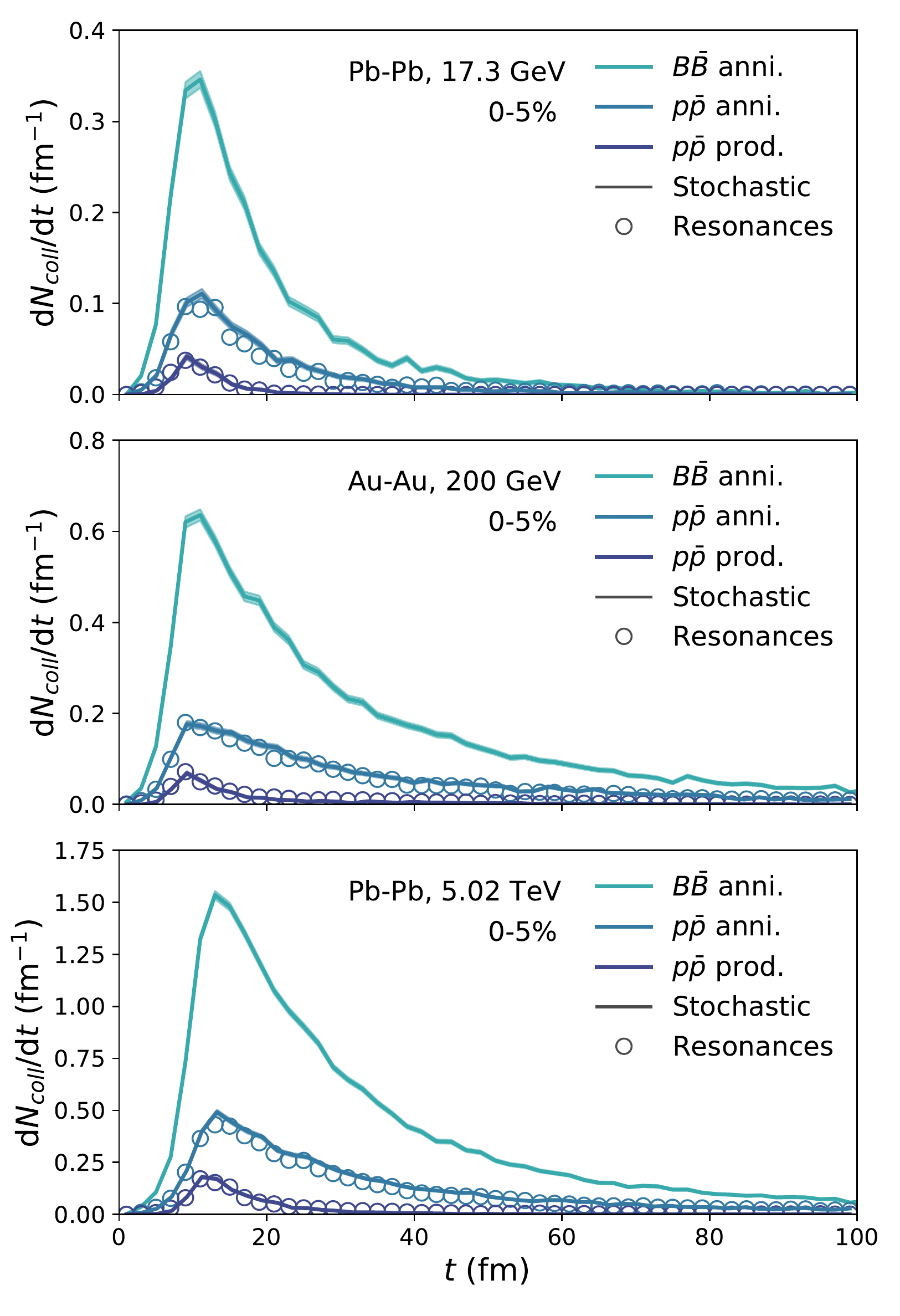}
	\caption{\label{fig:reaction-rates} Reaction rates for stochastic (lines) and resonance (symbols) treatment for different collision systems and energies: Pb-Pb at $\sqrt{s_{NN}}=17.3$ GeV (top), Au-Au at $\sqrt{s_{NN}}=200$ GeV (middle) and Pb-Pb at $\sqrt{s_{NN}}=5.02$ TeV (bottom). }
\end{figure}

In Fig.~\ref{fig:yields} one can see that, as expected from previous works, the inclusion of annihilations reduces the (anti-) proton yields \cite{Becattini:2012sq,*Steinheimer:2012rd}, and leads to a better agreement with experimental data. While the inclusion of detailed balance increases again the yield, we still observe a net decrease of the (anti-)proton yield. The effect of the $\ppbar$ regeneration can be grasped from the difference between the dashed and solid lines, which becomes more prominent for increasing collision energy, the ratio is referred to as 'regeneration factor' in the following. Full regeneration of the yield (see Ref.~\cite{Rapp:2000gy}) is not found for any of the systems. 

Our results show that 20-50$\%$ of proton-antiproton pairs lost to annihilation are regenerated, depending on the rapidity cuts.  We find that for the midrapidity region (see Fig.~\ref{fig:yields}), it is 40-50\% for all systems. For the full $4\pi$ multiplicity  20-30\% of the $\ppbar$ pairs lost to annihilation are regenerated. These results are compatible with previous works which employed a simplified scenario~\cite{Pan:2012ne}, as the authors also report a change of $20\%$ on the (4$\pi$) yield when including back-reactions. Furthermore, comparing to results from Ref.~\cite{Seifert:2018bwl}, which treats a more extensive set of $\BBbar$ reactions via detailed balance conserving 3-to-2 reactions, we find a similar net reduction of protons for $\snn = 200$ GeV. Conversely, our findings disagree with theirs at $\snn = 5.02$ TeV. We also checked the average transverse momentum of the (anti-)protons  and found negligible effects due to the inclusion of the back-reaction. 

Notice that while the difference between the $\rmP$ and $\pbar$ multiplicities fades away when increasing the collision energy (as the average $\mu_B$ approaches zero), we still find differences even at $\sqrt{s_{NN}}=5.02$ TeV. This difference can be traced back to a systematic positive (but small) value of $\mu_B$ at particlization that is reflected in a finite net baryon number in the system (as observed in the small difference between $\rmP$ and $\pbar$ yields at $t=0$ in Fig.~\ref{fig:reaction-rates}), which leads to the final $\rmP$ and $\pbar$ difference. This deficiency which does not match the experimental observation of equal yields at the highest beam energies can be alleviated by adjusting the initial state in the future. 

To gauge how strongly an intermediate resonance affects the results, we track the number of collisions, both total (Fig.~\ref{fig:Reactions}) and their time dependence (Fig.~\ref{fig:reaction-rates}). For the resonance treatment we choose to track only the $\ppbar\leftrightarrow h_1 \rho$ reaction as the annihilation/regeneration reaction, since the scattering partners of the intermediate reactions may mix into the medium, causing interrupted annihilations in non-equilibrium settings because of i.e. reabsorption of $\rho$ particles.
In Fig.~ \ref{fig:Reactions} the reader finds the quantification of the total number of reactions across three centrality classes, and a large range of energies. We also present two relevant ratios. First, the ratio of backward/forward reactions, which lies at around 15-20\%, stable in all systems. This number further confirms our agreement with Ref.~\cite{Pan:2012ne}.  Second, the ratio of $\ppbar$ annihilations to the number of (non-nucleon) baryon annihilations, which is also a constant value across all systems. Thanks to the stability of these numbers, we can use the two numbers to extrapolate our findings to the full spectrum of $\BBbar$ annihilation and regeneration.

In Fig.~\ref{fig:reaction-rates} it can be observed that both back-reaction treatments are consistent for the forward and backward reaction rate ($\ppbar$ annihilation and production), also when taking into account the time evolution. This is in contrast to previous studies for $d$ catalysis reactions at lower beam energies, where multi-step and multi-particle treatments revealed differences~\cite{Staudenmaier:2021lrg}. In fact, this excellent agreement is relatively surprising considering the dynamical differences of both the approaches, i.e.  intermediate state in-medium mixing. A possible explanation could be that a $\pi$-rich and longer-lived medium minimizes the effects of a slower reaction. So, while the stochastic treatment is more rigorous and flexible, it appears that the 5-to-2 reaction can be modeled in multiple steps without affecting the presented results for reaction rate (and resulting yield, see Fig.~\ref{fig:yields}).  Figures~\ref{fig:yields} and~\ref{fig:Reactions} are the main results of this letter, and can be summarized as the proof of consistence of the resonance and stochastic treatment and the quantification of  the $\ppbar$ regeneration in a fully microscopic description of the late stages.

\section{Conclusions}
\label{sec:conclusions}
In this work we have presented the first afterburner calculation employing detailed balance for proton-antiproton annihilation, namely the process $\ppbar \leftrightarrow 5 \pi$. We have found the backward/forward reaction ratio to be of 15-20\%, stable throughout a large range of collision energies and centralities.  This number is compatible with the regeneration factor found for $4\pi$ multiplicities in previous studies \cite{Pan:2012ne}.
This is also compatible with the trend we observe for the non-negligible  rapidity-dependent  regeneration factor in the (anti-)proton yields. For the full rapidity range a regeneration factor of 20-30\% is determined, while for the midrapidity slice, $|y|<0.5$, around 50$\%$ of the $\ppbar$ pairs lost to annihilation are regenerated.

This study supports the message that $\ppbar$ processes and the dynamical interplay of annihilation and regeneration does play a non-negligible role for the proton yield, and has to be accounted for when modeling the system. On the other hand, their effect on other observables, i.e. flow coefficients, are yet to be explored. Our results indicate that observing an equilibrated proton yield corresponding to a single temperature and chemical potential at the chemical freeze-out is not possible, even when taking detailed balanced multi-particle reactions into account. The expansion of the system superseeds the reaction probabilities and non-equilibrium effects are important. Furthermore, our study presents the opportunity to use the findings for the specific $\ppbar$ reaction to estimate the effect that general $B\bar{B}$ annihilation and back-reaction have on the yields of the remaining baryons. 

In the future, quantitative extraction of properties of the quark-gluon plasma created in heavy-ion collisions that are based on proton yields as a basic bulk observable have to take into account the significant effects of annihilation and regeneration in the late stage hadronic evolution. This might have an effect on the other parameters of the calculation, such as the switching transition criterion. The stochastic treatment can be further extended to include general baryon-antibaryon annihilation, as well as the back-reactions to other $2\rightarrow n$ processes.  

\section{Acknowledgements}
We thank Tim Neidig for providing the solution of the analytical rate equations in Ref.~\cite{TimMaster}, included in our Fig.~\ref{fig:box}. This project was  supported by  the Deutsche Forschungsgemeinschaft (DFG, German Research Foundation) – Project number 315477589 – TRR 211. A.S. acknowledges support by the Stiftung Polytechnische Gesellschaft Frankfurt am Main as well as the GSI F\&E program. Computational resources have been provided by the Center for Scientific Computing (CSC) at the Goethe-University of Frankfurt and the GreenCube at GSI.

\bibliographystyle{apsrev4-1}
\bibliography{References}

\begin{thebibliography}{35}%
\makeatletter
\providecommand \@ifxundefined [1]{%
 \@ifx{#1\undefined}
}%
\providecommand \@ifnum [1]{%
 \ifnum #1\expandafter \@firstoftwo
 \else \expandafter \@secondoftwo
 \fi
}%
\providecommand \@ifx [1]{%
 \ifx #1\expandafter \@firstoftwo
 \else \expandafter \@secondoftwo
 \fi
}%
\providecommand \natexlab [1]{#1}%
\providecommand \enquote  [1]{``#1''}%
\providecommand \bibnamefont  [1]{#1}%
\providecommand \bibfnamefont [1]{#1}%
\providecommand \citenamefont [1]{#1}%
\providecommand \href@noop [0]{\@secondoftwo}%
\providecommand \href [0]{\begingroup \@sanitize@url \@href}%
\providecommand \@href[1]{\@@startlink{#1}\@@href}%
\providecommand \@@href[1]{\endgroup#1\@@endlink}%
\providecommand \@sanitize@url [0]{\catcode `\\12\catcode `\$12\catcode
  `\&12\catcode `\#12\catcode `\^12\catcode `\_12\catcode `\%12\relax}%
\providecommand \@@startlink[1]{}%
\providecommand \@@endlink[0]{}%
\providecommand \url  [0]{\begingroup\@sanitize@url \@url }%
\providecommand \@url [1]{\endgroup\@href {#1}{\urlprefix }}%
\providecommand \urlprefix  [0]{URL }%
\providecommand \Eprint [0]{\href }%
\providecommand \doibase [0]{http://dx.doi.org/}%
\providecommand \selectlanguage [0]{\@gobble}%
\providecommand \bibinfo  [0]{\@secondoftwo}%
\providecommand \bibfield  [0]{\@secondoftwo}%
\providecommand \translation [1]{[#1]}%
\providecommand \BibitemOpen [0]{}%
\providecommand \bibitemStop [0]{}%
\providecommand \bibitemNoStop [0]{.\EOS\space}%
\providecommand \EOS [0]{\spacefactor3000\relax}%
\providecommand \BibitemShut  [1]{\csname bibitem#1\endcsname}%
\let\auto@bib@innerbib\@empty
\bibitem [{\citenamefont {Andronic}\ \emph {et~al.}(2013)\citenamefont
  {Andronic}, \citenamefont {Braun-Munzinger}, \citenamefont {Redlich},\ and\
  \citenamefont {Stachel}}]{Andronic:2012dm}%
  \BibitemOpen
  \bibfield  {author} {\bibinfo {author} {\bibfnamefont {A.}~\bibnamefont
  {Andronic}}, \bibinfo {author} {\bibfnamefont {P.}~\bibnamefont
  {Braun-Munzinger}}, \bibinfo {author} {\bibfnamefont {K.}~\bibnamefont
  {Redlich}}, \ and\ \bibinfo {author} {\bibfnamefont {J.}~\bibnamefont
  {Stachel}},\ }\href {\doibase 10.1016/j.nuclphysa.2013.02.070} {\bibfield
  {journal} {\bibinfo  {journal} {Nucl. Phys. A}\ }\textbf {\bibinfo {volume}
  {904-905}},\ \bibinfo {pages} {535c} (\bibinfo {year} {2013})},\ \Eprint
  {http://arxiv.org/abs/1210.7724} {arXiv:1210.7724 [nucl-th]} \BibitemShut
  {NoStop}%
\bibitem [{\citenamefont {Andronic}\ \emph {et~al.}(2018)\citenamefont
  {Andronic}, \citenamefont {Braun-Munzinger}, \citenamefont {Redlich},\ and\
  \citenamefont {Stachel}}]{Andronic:2017pug}%
  \BibitemOpen
  \bibfield  {author} {\bibinfo {author} {\bibfnamefont {A.}~\bibnamefont
  {Andronic}}, \bibinfo {author} {\bibfnamefont {P.}~\bibnamefont
  {Braun-Munzinger}}, \bibinfo {author} {\bibfnamefont {K.}~\bibnamefont
  {Redlich}}, \ and\ \bibinfo {author} {\bibfnamefont {J.}~\bibnamefont
  {Stachel}},\ }\href {\doibase 10.1038/s41586-018-0491-6} {\bibfield
  {journal} {\bibinfo  {journal} {Nature}\ }\textbf {\bibinfo {volume} {561}},\
  \bibinfo {pages} {321} (\bibinfo {year} {2018})},\ \Eprint
  {http://arxiv.org/abs/1710.09425} {arXiv:1710.09425 [nucl-th]} \BibitemShut
  {NoStop}%
\bibitem [{\citenamefont {Stachel}\ \emph {et~al.}(2014)\citenamefont
  {Stachel}, \citenamefont {Andronic}, \citenamefont {Braun-Munzinger},\ and\
  \citenamefont {Redlich}}]{Stachel:2013zma}%
  \BibitemOpen
  \bibfield  {author} {\bibinfo {author} {\bibfnamefont {J.}~\bibnamefont
  {Stachel}}, \bibinfo {author} {\bibfnamefont {A.}~\bibnamefont {Andronic}},
  \bibinfo {author} {\bibfnamefont {P.}~\bibnamefont {Braun-Munzinger}}, \ and\
  \bibinfo {author} {\bibfnamefont {K.}~\bibnamefont {Redlich}},\ }\href
  {\doibase 10.1088/1742-6596/509/1/012019} {\bibfield  {journal} {\bibinfo
  {journal} {J. Phys. Conf. Ser.}\ }\textbf {\bibinfo {volume} {509}},\
  \bibinfo {pages} {012019} (\bibinfo {year} {2014})},\ \Eprint
  {http://arxiv.org/abs/1311.4662} {arXiv:1311.4662 [nucl-th]} \BibitemShut
  {NoStop}%
\bibitem [{\citenamefont {Andronic}\ \emph {et~al.}(2019)\citenamefont
  {Andronic}, \citenamefont {Braun-Munzinger}, \citenamefont {Friman},
  \citenamefont {Lo}, \citenamefont {Redlich},\ and\ \citenamefont
  {Stachel}}]{Andronic:2018qqt}%
  \BibitemOpen
  \bibfield  {author} {\bibinfo {author} {\bibfnamefont {A.}~\bibnamefont
  {Andronic}}, \bibinfo {author} {\bibfnamefont {P.}~\bibnamefont
  {Braun-Munzinger}}, \bibinfo {author} {\bibfnamefont {B.}~\bibnamefont
  {Friman}}, \bibinfo {author} {\bibfnamefont {P.~M.}\ \bibnamefont {Lo}},
  \bibinfo {author} {\bibfnamefont {K.}~\bibnamefont {Redlich}}, \ and\
  \bibinfo {author} {\bibfnamefont {J.}~\bibnamefont {Stachel}},\ }\href
  {\doibase 10.1016/j.physletb.2019.03.052} {\bibfield  {journal} {\bibinfo
  {journal} {Phys. Lett. B}\ }\textbf {\bibinfo {volume} {792}},\ \bibinfo
  {pages} {304} (\bibinfo {year} {2019})},\ \Eprint
  {http://arxiv.org/abs/1808.03102} {arXiv:1808.03102 [hep-ph]} \BibitemShut
  {NoStop}%
\bibitem [{\citenamefont {Becattini}\ \emph {et~al.}(2012)\citenamefont
  {Becattini}, \citenamefont {Bleicher}, \citenamefont {Kollegger},
  \citenamefont {Mitrovski}, \citenamefont {Schuster},\ and\ \citenamefont
  {Stock}}]{Becattini:2012sq}%
  \BibitemOpen
  \bibfield  {author} {\bibinfo {author} {\bibfnamefont {F.}~\bibnamefont
  {Becattini}}, \bibinfo {author} {\bibfnamefont {M.}~\bibnamefont {Bleicher}},
  \bibinfo {author} {\bibfnamefont {T.}~\bibnamefont {Kollegger}}, \bibinfo
  {author} {\bibfnamefont {M.}~\bibnamefont {Mitrovski}}, \bibinfo {author}
  {\bibfnamefont {T.}~\bibnamefont {Schuster}}, \ and\ \bibinfo {author}
  {\bibfnamefont {R.}~\bibnamefont {Stock}},\ }\href {\doibase
  10.1103/PhysRevC.85.044921} {\bibfield  {journal} {\bibinfo  {journal} {Phys.
  Rev. C}\ }\textbf {\bibinfo {volume} {85}},\ \bibinfo {pages} {044921}
  (\bibinfo {year} {2012})},\ \Eprint {http://arxiv.org/abs/1201.6349}
  {arXiv:1201.6349 [nucl-th]} \BibitemShut {NoStop}%
\bibitem [{\citenamefont {Steinheimer}\ \emph {et~al.}(2013)\citenamefont
  {Steinheimer}, \citenamefont {Aichelin},\ and\ \citenamefont
  {Bleicher}}]{Steinheimer:2012rd}%
  \BibitemOpen
  \bibfield  {author} {\bibinfo {author} {\bibfnamefont {J.}~\bibnamefont
  {Steinheimer}}, \bibinfo {author} {\bibfnamefont {J.}~\bibnamefont
  {Aichelin}}, \ and\ \bibinfo {author} {\bibfnamefont {M.}~\bibnamefont
  {Bleicher}},\ }\href {\doibase 10.1103/PhysRevLett.110.042501} {\bibfield
  {journal} {\bibinfo  {journal} {Phys. Rev. Lett.}\ }\textbf {\bibinfo
  {volume} {110}},\ \bibinfo {pages} {042501} (\bibinfo {year} {2013})},\
  \Eprint {http://arxiv.org/abs/1203.5302} {arXiv:1203.5302 [nucl-th]}
  \BibitemShut {NoStop}%
\bibitem [{\citenamefont {Karpenko}\ \emph {et~al.}(2013)\citenamefont
  {Karpenko}, \citenamefont {Sinyukov},\ and\ \citenamefont
  {Werner}}]{Karpenko:2012yf}%
  \BibitemOpen
  \bibfield  {author} {\bibinfo {author} {\bibfnamefont {I.~A.}\ \bibnamefont
  {Karpenko}}, \bibinfo {author} {\bibfnamefont {Y.~M.}\ \bibnamefont
  {Sinyukov}}, \ and\ \bibinfo {author} {\bibfnamefont {K.}~\bibnamefont
  {Werner}},\ }\href {\doibase 10.1103/PhysRevC.87.024914} {\bibfield
  {journal} {\bibinfo  {journal} {Phys. Rev. C}\ }\textbf {\bibinfo {volume}
  {87}},\ \bibinfo {pages} {024914} (\bibinfo {year} {2013})},\ \Eprint
  {http://arxiv.org/abs/1204.5351} {arXiv:1204.5351 [nucl-th]} \BibitemShut
  {NoStop}%
\bibitem [{\citenamefont {Rapp}\ and\ \citenamefont
  {Shuryak}(2001)}]{Rapp:2000gy}%
  \BibitemOpen
  \bibfield  {author} {\bibinfo {author} {\bibfnamefont {R.}~\bibnamefont
  {Rapp}}\ and\ \bibinfo {author} {\bibfnamefont {E.~V.}\ \bibnamefont
  {Shuryak}},\ }\href {\doibase 10.1103/PhysRevLett.86.2980} {\bibfield
  {journal} {\bibinfo  {journal} {Phys. Rev. Lett.}\ }\textbf {\bibinfo
  {volume} {86}},\ \bibinfo {pages} {2980} (\bibinfo {year} {2001})},\ \Eprint
  {http://arxiv.org/abs/hep-ph/0008326} {arXiv:hep-ph/0008326} \BibitemShut
  {NoStop}%
\bibitem [{\citenamefont {Pan}\ and\ \citenamefont {Pratt}(2014)}]{Pan:2012ne}%
  \BibitemOpen
  \bibfield  {author} {\bibinfo {author} {\bibfnamefont {Y.}~\bibnamefont
  {Pan}}\ and\ \bibinfo {author} {\bibfnamefont {S.}~\bibnamefont {Pratt}},\
  }\href@noop {} {\enquote {\bibinfo {title} {Baryon annihilation and
  regeneration in heavy ion collisions},}\ } (\bibinfo {year} {2014}),\ \Eprint
  {http://arxiv.org/abs/1210.1577} {arXiv:1210.1577 [nucl-th]} \BibitemShut
  {NoStop}%
\bibitem [{\citenamefont {Ryu}\ \emph {et~al.}(2018)\citenamefont {Ryu},
  \citenamefont {Paquet}, \citenamefont {Shen}, \citenamefont {Denicol},
  \citenamefont {Schenke}, \citenamefont {Jeon},\ and\ \citenamefont
  {Gale}}]{Ryu:2017qzn}%
  \BibitemOpen
  \bibfield  {author} {\bibinfo {author} {\bibfnamefont {S.}~\bibnamefont
  {Ryu}}, \bibinfo {author} {\bibfnamefont {J.-F.}\ \bibnamefont {Paquet}},
  \bibinfo {author} {\bibfnamefont {C.}~\bibnamefont {Shen}}, \bibinfo {author}
  {\bibfnamefont {G.}~\bibnamefont {Denicol}}, \bibinfo {author} {\bibfnamefont
  {B.}~\bibnamefont {Schenke}}, \bibinfo {author} {\bibfnamefont
  {S.}~\bibnamefont {Jeon}}, \ and\ \bibinfo {author} {\bibfnamefont
  {C.}~\bibnamefont {Gale}},\ }\href {\doibase 10.1103/PhysRevC.97.034910}
  {\bibfield  {journal} {\bibinfo  {journal} {Phys. Rev. C}\ }\textbf {\bibinfo
  {volume} {97}},\ \bibinfo {pages} {034910} (\bibinfo {year} {2018})},\
  \Eprint {http://arxiv.org/abs/1704.04216} {arXiv:1704.04216 [nucl-th]}
  \BibitemShut {NoStop}%
\bibitem [{\citenamefont {Everett}\ \emph
  {et~al.}(2021{\natexlab{a}})\citenamefont {Everett} \emph
  {et~al.}}]{JETSCAPE:2020mzn}%
  \BibitemOpen
  \bibfield  {author} {\bibinfo {author} {\bibfnamefont {D.}~\bibnamefont
  {Everett}} \emph {et~al.} (\bibinfo {collaboration} {JETSCAPE}),\ }\href
  {\doibase 10.1103/PhysRevC.103.054904} {\bibfield  {journal} {\bibinfo
  {journal} {Phys. Rev. C}\ }\textbf {\bibinfo {volume} {103}},\ \bibinfo
  {pages} {054904} (\bibinfo {year} {2021}{\natexlab{a}})},\ \Eprint
  {http://arxiv.org/abs/2011.01430} {arXiv:2011.01430 [hep-ph]} \BibitemShut
  {NoStop}%
\bibitem [{\citenamefont {Everett}\ \emph
  {et~al.}(2021{\natexlab{b}})\citenamefont {Everett} \emph
  {et~al.}}]{JETSCAPE:2020shq}%
  \BibitemOpen
  \bibfield  {author} {\bibinfo {author} {\bibfnamefont {D.}~\bibnamefont
  {Everett}} \emph {et~al.} (\bibinfo {collaboration} {JETSCAPE}),\ }\href
  {\doibase 10.1103/PhysRevLett.126.242301} {\bibfield  {journal} {\bibinfo
  {journal} {Phys. Rev. Lett.}\ }\textbf {\bibinfo {volume} {126}},\ \bibinfo
  {pages} {242301} (\bibinfo {year} {2021}{\natexlab{b}})},\ \Eprint
  {http://arxiv.org/abs/2010.03928} {arXiv:2010.03928 [hep-ph]} \BibitemShut
  {NoStop}%
\bibitem [{\citenamefont {Staudenmaier}\ \emph {et~al.}(2021)\citenamefont
  {Staudenmaier}, \citenamefont {Oliinychenko}, \citenamefont {Torres-Rincon},\
  and\ \citenamefont {Elfner}}]{Staudenmaier:2021lrg}%
  \BibitemOpen
  \bibfield  {author} {\bibinfo {author} {\bibfnamefont {J.}~\bibnamefont
  {Staudenmaier}}, \bibinfo {author} {\bibfnamefont {D.}~\bibnamefont
  {Oliinychenko}}, \bibinfo {author} {\bibfnamefont {J.~M.}\ \bibnamefont
  {Torres-Rincon}}, \ and\ \bibinfo {author} {\bibfnamefont {H.}~\bibnamefont
  {Elfner}},\ }\href@noop {} {\  (\bibinfo {year} {2021})},\ \Eprint
  {http://arxiv.org/abs/2106.14287} {arXiv:2106.14287 [hep-ph]} \BibitemShut
  {NoStop}%
\bibitem [{\citenamefont {Bass}\ \emph {et~al.}(1998)\citenamefont {Bass} \emph
  {et~al.}}]{Bass:1998ca}%
  \BibitemOpen
  \bibfield  {author} {\bibinfo {author} {\bibfnamefont {S.~A.}\ \bibnamefont
  {Bass}} \emph {et~al.},\ }\href {\doibase 10.1016/S0146-6410(98)00058-1}
  {\bibfield  {journal} {\bibinfo  {journal} {Prog. Part. Nucl. Phys.}\
  }\textbf {\bibinfo {volume} {41}},\ \bibinfo {pages} {255} (\bibinfo {year}
  {1998})},\ \Eprint {http://arxiv.org/abs/nucl-th/9803035}
  {arXiv:nucl-th/9803035} \BibitemShut {NoStop}%
\bibitem [{\citenamefont {Weil}\ \emph {et~al.}(2016)\citenamefont {Weil} \emph
  {et~al.}}]{Weil:2016zrk}%
  \BibitemOpen
  \bibfield  {author} {\bibinfo {author} {\bibfnamefont {J.}~\bibnamefont
  {Weil}} \emph {et~al.},\ }\href {\doibase 10.1103/PhysRevC.94.054905}
  {\bibfield  {journal} {\bibinfo  {journal} {Phys. Rev. C}\ }\textbf {\bibinfo
  {volume} {94}},\ \bibinfo {pages} {054905} (\bibinfo {year} {2016})},\
  \Eprint {http://arxiv.org/abs/1606.06642} {arXiv:1606.06642 [nucl-th]}
  \BibitemShut {NoStop}%
\bibitem [{\citenamefont {Cassing}(2002)}]{Cassing:2001ds}%
  \BibitemOpen
  \bibfield  {author} {\bibinfo {author} {\bibfnamefont {W.}~\bibnamefont
  {Cassing}},\ }\href {\doibase 10.1016/S0375-9474(01)01322-7} {\bibfield
  {journal} {\bibinfo  {journal} {Nucl. Phys. A}\ }\textbf {\bibinfo {volume}
  {700}},\ \bibinfo {pages} {618} (\bibinfo {year} {2002})},\ \Eprint
  {http://arxiv.org/abs/nucl-th/0105069} {arXiv:nucl-th/0105069} \BibitemShut
  {NoStop}%
\bibitem [{\citenamefont {Seifert}\ and\ \citenamefont
  {Cassing}(2018{\natexlab{a}})}]{Seifert:2017oyb}%
  \BibitemOpen
  \bibfield  {author} {\bibinfo {author} {\bibfnamefont {E.}~\bibnamefont
  {Seifert}}\ and\ \bibinfo {author} {\bibfnamefont {W.}~\bibnamefont
  {Cassing}},\ }\href {\doibase 10.1103/PhysRevC.97.024913} {\bibfield
  {journal} {\bibinfo  {journal} {Phys. Rev. C}\ }\textbf {\bibinfo {volume}
  {97}},\ \bibinfo {pages} {024913} (\bibinfo {year} {2018}{\natexlab{a}})},\
  \Eprint {http://arxiv.org/abs/1710.00665} {arXiv:1710.00665 [hep-ph]}
  \BibitemShut {NoStop}%
\bibitem [{\citenamefont {Seifert}\ and\ \citenamefont
  {Cassing}(2018{\natexlab{b}})}]{Seifert:2018bwl}%
  \BibitemOpen
  \bibfield  {author} {\bibinfo {author} {\bibfnamefont {E.}~\bibnamefont
  {Seifert}}\ and\ \bibinfo {author} {\bibfnamefont {W.}~\bibnamefont
  {Cassing}},\ }\href {\doibase 10.1103/PhysRevC.97.044907} {\bibfield
  {journal} {\bibinfo  {journal} {Phys. Rev. C}\ }\textbf {\bibinfo {volume}
  {97}},\ \bibinfo {pages} {044907} (\bibinfo {year} {2018}{\natexlab{b}})},\
  \Eprint {http://arxiv.org/abs/1801.07557} {arXiv:1801.07557 [hep-ph]}
  \BibitemShut {NoStop}%
\bibitem [{\citenamefont {Neidig}(2020)}]{TimMaster}%
  \BibitemOpen
  \bibfield  {author} {\bibinfo {author} {\bibfnamefont {T.}~\bibnamefont
  {Neidig}},\ }\emph {\bibinfo {title} {Production of light nuclei in
  ultra-relativistic heavy-ion collisions via a network of rate equations}},\
  \href@noop {} {Master's thesis},\ \bibinfo  {school} {Goethe Universit\"at
  Frankfurt} (\bibinfo {year} {2020})\BibitemShut {NoStop}%
\bibitem [{SVH()}]{SVH}%
  \BibitemOpen
  \href@noop {} {\enquote {\bibinfo {title} {Smash-vhlle-hybrid},}\ }\bibinfo
  {howpublished}
  {https://github.com/smash-transport/smash-vhlle-hybrid}\BibitemShut {NoStop}%
\bibitem [{\citenamefont {Oliinychenko}\ \emph {et~al.}(2020)\citenamefont
  {Oliinychenko}, \citenamefont {Steinberg}, \citenamefont {Weil},
  \citenamefont {Staudenmaier}, \citenamefont {Kretz}, \citenamefont
  {Schäfer}, \citenamefont {(Petersen)}, \citenamefont {Ryu}, \citenamefont
  {Rothermel}, \citenamefont {Mohs}, \citenamefont {Li}, \citenamefont
  {Sorensen}, \citenamefont {Mitrovic}, \citenamefont {Pang}, \citenamefont
  {Hammelmann}, \citenamefont {Goldschmidt}, \citenamefont {Mayer},
  \citenamefont {Garcia-Montero}, \citenamefont {Kübler},\ and\ \citenamefont
  {Nikita}}]{SMASH:DOI}%
  \BibitemOpen
  \bibfield  {author} {\bibinfo {author} {\bibfnamefont {D.}~\bibnamefont
  {Oliinychenko}}, \bibinfo {author} {\bibfnamefont {V.}~\bibnamefont
  {Steinberg}}, \bibinfo {author} {\bibfnamefont {J.}~\bibnamefont {Weil}},
  \bibinfo {author} {\bibfnamefont {J.}~\bibnamefont {Staudenmaier}}, \bibinfo
  {author} {\bibfnamefont {M.}~\bibnamefont {Kretz}}, \bibinfo {author}
  {\bibfnamefont {A.}~\bibnamefont {Schäfer}}, \bibinfo {author}
  {\bibfnamefont {H.~E.}\ \bibnamefont {(Petersen)}}, \bibinfo {author}
  {\bibfnamefont {S.}~\bibnamefont {Ryu}}, \bibinfo {author} {\bibfnamefont
  {J.}~\bibnamefont {Rothermel}}, \bibinfo {author} {\bibfnamefont
  {J.}~\bibnamefont {Mohs}}, \bibinfo {author} {\bibfnamefont {F.}~\bibnamefont
  {Li}}, \bibinfo {author} {\bibfnamefont {A.}~\bibnamefont {Sorensen}},
  \bibinfo {author} {\bibfnamefont {D.}~\bibnamefont {Mitrovic}}, \bibinfo
  {author} {\bibfnamefont {L.}~\bibnamefont {Pang}}, \bibinfo {author}
  {\bibfnamefont {J.}~\bibnamefont {Hammelmann}}, \bibinfo {author}
  {\bibfnamefont {A.}~\bibnamefont {Goldschmidt}}, \bibinfo {author}
  {\bibfnamefont {M.}~\bibnamefont {Mayer}}, \bibinfo {author} {\bibfnamefont
  {O.}~\bibnamefont {Garcia-Montero}}, \bibinfo {author} {\bibfnamefont
  {N.}~\bibnamefont {Kübler}}, \ and\ \bibinfo {author} {\bibnamefont
  {Nikita}},\ }\href {\doibase 10.5281/zenodo.4336358} {\enquote {\bibinfo
  {title} {smash-transport/smash: Smash-2.0},}\ } (\bibinfo {year}
  {2020})\BibitemShut {NoStop}%
\bibitem [{\citenamefont {Karpenko}\ \emph {et~al.}(2014)\citenamefont
  {Karpenko}, \citenamefont {Huovinen},\ and\ \citenamefont
  {Bleicher}}]{Karpenko:2013wva}%
  \BibitemOpen
  \bibfield  {author} {\bibinfo {author} {\bibfnamefont {I.}~\bibnamefont
  {Karpenko}}, \bibinfo {author} {\bibfnamefont {P.}~\bibnamefont {Huovinen}},
  \ and\ \bibinfo {author} {\bibfnamefont {M.}~\bibnamefont {Bleicher}},\
  }\href {\doibase 10.1016/j.cpc.2014.07.010} {\bibfield  {journal} {\bibinfo
  {journal} {Comput. Phys. Commun.}\ }\textbf {\bibinfo {volume} {185}},\
  \bibinfo {pages} {3016} (\bibinfo {year} {2014})},\ \Eprint
  {http://arxiv.org/abs/1312.4160} {arXiv:1312.4160 [nucl-th]} \BibitemShut
  {NoStop}%
\bibitem [{\citenamefont {Karpenko}\ \emph {et~al.}(2015)\citenamefont
  {Karpenko}, \citenamefont {Huovinen}, \citenamefont {Petersen},\ and\
  \citenamefont {Bleicher}}]{Karpenko:2015xea}%
  \BibitemOpen
  \bibfield  {author} {\bibinfo {author} {\bibfnamefont {I.}~\bibnamefont
  {Karpenko}}, \bibinfo {author} {\bibfnamefont {P.}~\bibnamefont {Huovinen}},
  \bibinfo {author} {\bibfnamefont {H.}~\bibnamefont {Petersen}}, \ and\
  \bibinfo {author} {\bibfnamefont {M.}~\bibnamefont {Bleicher}},\ }\href
  {\doibase 10.1103/PhysRevC.91.064901} {\bibfield  {journal} {\bibinfo
  {journal} {Phys. Rev. C}\ }\textbf {\bibinfo {volume} {91}},\ \bibinfo
  {pages} {064901} (\bibinfo {year} {2015})},\ \Eprint
  {http://arxiv.org/abs/1502.01978} {arXiv:1502.01978 [nucl-th]} \BibitemShut
  {NoStop}%
\bibitem [{SHS()}]{SHS}%
  \BibitemOpen
  \href@noop {} {\enquote {\bibinfo {title} {Smash-hadron-sampler},}\ }\bibinfo
  {howpublished}
  {https://github.com/smash-transport/smash-hadron-sampler}\BibitemShut
  {NoStop}%
\bibitem [{\citenamefont {Steinheimer}\ \emph {et~al.}(2011)\citenamefont
  {Steinheimer}, \citenamefont {Schramm},\ and\ \citenamefont
  {Stocker}}]{Steinheimer:2010ib}%
  \BibitemOpen
  \bibfield  {author} {\bibinfo {author} {\bibfnamefont {J.}~\bibnamefont
  {Steinheimer}}, \bibinfo {author} {\bibfnamefont {S.}~\bibnamefont
  {Schramm}}, \ and\ \bibinfo {author} {\bibfnamefont {H.}~\bibnamefont
  {Stocker}},\ }\href {\doibase 10.1088/0954-3899/38/3/035001} {\bibfield
  {journal} {\bibinfo  {journal} {J. Phys. G}\ }\textbf {\bibinfo {volume}
  {38}},\ \bibinfo {pages} {035001} (\bibinfo {year} {2011})},\ \Eprint
  {http://arxiv.org/abs/1009.5239} {arXiv:1009.5239 [hep-ph]} \BibitemShut
  {NoStop}%
\bibitem [{\citenamefont {Huovinen}\ and\ \citenamefont
  {Petersen}(2012)}]{Huovinen:2012is}%
  \BibitemOpen
  \bibfield  {author} {\bibinfo {author} {\bibfnamefont {P.}~\bibnamefont
  {Huovinen}}\ and\ \bibinfo {author} {\bibfnamefont {H.}~\bibnamefont
  {Petersen}},\ }\href {\doibase 10.1140/epja/i2012-12171-9} {\bibfield
  {journal} {\bibinfo  {journal} {Eur. Phys. J. A}\ }\textbf {\bibinfo {volume}
  {48}},\ \bibinfo {pages} {171} (\bibinfo {year} {2012})},\ \Eprint
  {http://arxiv.org/abs/1206.3371} {arXiv:1206.3371 [nucl-th]} \BibitemShut
  {NoStop}%
\bibitem [{\citenamefont {Mohs}\ \emph {et~al.}(2020)\citenamefont {Mohs},
  \citenamefont {Ryu},\ and\ \citenamefont {Elfner}}]{Mohs:2019iee}%
  \BibitemOpen
  \bibfield  {author} {\bibinfo {author} {\bibfnamefont {J.}~\bibnamefont
  {Mohs}}, \bibinfo {author} {\bibfnamefont {S.}~\bibnamefont {Ryu}}, \ and\
  \bibinfo {author} {\bibfnamefont {H.}~\bibnamefont {Elfner}},\ }\href
  {\doibase 10.1088/1361-6471/ab7bd1} {\bibfield  {journal} {\bibinfo
  {journal} {J. Phys. G}\ }\textbf {\bibinfo {volume} {47}},\ \bibinfo {pages}
  {065101} (\bibinfo {year} {2020})},\ \Eprint
  {http://arxiv.org/abs/1909.05586} {arXiv:1909.05586 [nucl-th]} \BibitemShut
  {NoStop}%
\bibitem [{\citenamefont {Demir}(2010)}]{demirphd}%
  \BibitemOpen
  \bibfield  {author} {\bibinfo {author} {\bibfnamefont {N.}~\bibnamefont
  {Demir}},\ }\emph {\bibinfo {title} {Extraction of Hot QCD Matter Transport
  Coefficients utilizing Microscopic Transport Theory. Dissertation, Duke
  University.}},\ \href@noop {} {Ph.D. thesis},\ \bibinfo  {school} {Duke
  University} (\bibinfo {year} {2010})\BibitemShut {NoStop}%
\bibitem [{\citenamefont {Rose}\ \emph {et~al.}(2018)\citenamefont {Rose},
  \citenamefont {Torres-Rincon}, \citenamefont {Sch\"afer}, \citenamefont
  {Oliinychenko},\ and\ \citenamefont {Petersen}}]{Rose:2017bjz}%
  \BibitemOpen
  \bibfield  {author} {\bibinfo {author} {\bibfnamefont {J.~B.}\ \bibnamefont
  {Rose}}, \bibinfo {author} {\bibfnamefont {J.~M.}\ \bibnamefont
  {Torres-Rincon}}, \bibinfo {author} {\bibfnamefont {A.}~\bibnamefont
  {Sch\"afer}}, \bibinfo {author} {\bibfnamefont {D.~R.}\ \bibnamefont
  {Oliinychenko}}, \ and\ \bibinfo {author} {\bibfnamefont {H.}~\bibnamefont
  {Petersen}},\ }\href {\doibase 10.1103/PhysRevC.97.055204} {\bibfield
  {journal} {\bibinfo  {journal} {Phys. Rev. C}\ }\textbf {\bibinfo {volume}
  {97}},\ \bibinfo {pages} {055204} (\bibinfo {year} {2018})},\ \Eprint
  {http://arxiv.org/abs/1709.03826} {arXiv:1709.03826 [nucl-th]} \BibitemShut
  {NoStop}%
\bibitem [{\citenamefont {Koch}\ \emph {et~al.}(1986)\citenamefont {Koch},
  \citenamefont {Muller},\ and\ \citenamefont {Rafelski}}]{Koch:1986ud}%
  \BibitemOpen
  \bibfield  {author} {\bibinfo {author} {\bibfnamefont {P.}~\bibnamefont
  {Koch}}, \bibinfo {author} {\bibfnamefont {B.}~\bibnamefont {Muller}}, \ and\
  \bibinfo {author} {\bibfnamefont {J.}~\bibnamefont {Rafelski}},\ }\href
  {\doibase 10.1016/0370-1573(86)90096-7} {\bibfield  {journal} {\bibinfo
  {journal} {Phys. Rept.}\ }\textbf {\bibinfo {volume} {142}},\ \bibinfo
  {pages} {167} (\bibinfo {year} {1986})}\BibitemShut {NoStop}%
\bibitem [{\citenamefont {Dover}\ \emph {et~al.}(1992)\citenamefont {Dover},
  \citenamefont {Gutsche}, \citenamefont {Maruyama},\ and\ \citenamefont
  {Faessler}}]{Dover:1992vj}%
  \BibitemOpen
  \bibfield  {author} {\bibinfo {author} {\bibfnamefont {C.~B.}\ \bibnamefont
  {Dover}}, \bibinfo {author} {\bibfnamefont {T.}~\bibnamefont {Gutsche}},
  \bibinfo {author} {\bibfnamefont {M.}~\bibnamefont {Maruyama}}, \ and\
  \bibinfo {author} {\bibfnamefont {A.}~\bibnamefont {Faessler}},\ }\href
  {\doibase 10.1016/0146-6410(92)90004-L} {\bibfield  {journal} {\bibinfo
  {journal} {Prog. Part. Nucl. Phys.}\ }\textbf {\bibinfo {volume} {29}},\
  \bibinfo {pages} {87} (\bibinfo {year} {1992})}\BibitemShut {NoStop}%
\bibitem [{\citenamefont {Acharya}\ \emph {et~al.}(2020)\citenamefont {Acharya}
  \emph {et~al.}}]{ALICE:2019hno}%
  \BibitemOpen
  \bibfield  {author} {\bibinfo {author} {\bibfnamefont {S.}~\bibnamefont
  {Acharya}} \emph {et~al.} (\bibinfo {collaboration} {ALICE}),\ }\href
  {\doibase 10.1103/PhysRevC.101.044907} {\bibfield  {journal} {\bibinfo
  {journal} {Phys. Rev. C}\ }\textbf {\bibinfo {volume} {101}},\ \bibinfo
  {pages} {044907} (\bibinfo {year} {2020})},\ \Eprint
  {http://arxiv.org/abs/1910.07678} {arXiv:1910.07678 [nucl-ex]} \BibitemShut
  {NoStop}%
\bibitem [{\citenamefont {Adler}\ \emph {et~al.}(2004)\citenamefont {Adler}
  \emph {et~al.}}]{PHENIX:2003iij}%
  \BibitemOpen
  \bibfield  {author} {\bibinfo {author} {\bibfnamefont {S.~S.}\ \bibnamefont
  {Adler}} \emph {et~al.} (\bibinfo {collaboration} {PHENIX}),\ }\href
  {\doibase 10.1103/PhysRevC.69.034909} {\bibfield  {journal} {\bibinfo
  {journal} {Phys. Rev. C}\ }\textbf {\bibinfo {volume} {69}},\ \bibinfo
  {pages} {034909} (\bibinfo {year} {2004})},\ \Eprint
  {http://arxiv.org/abs/nucl-ex/0307022} {arXiv:nucl-ex/0307022} \BibitemShut
  {NoStop}%
\bibitem [{\citenamefont {Becattini}\ \emph {et~al.}(2001)\citenamefont
  {Becattini}, \citenamefont {Cleymans}, \citenamefont {Keranen}, \citenamefont
  {Suhonen},\ and\ \citenamefont {Redlich}}]{Becattini:2000jw}%
  \BibitemOpen
  \bibfield  {author} {\bibinfo {author} {\bibfnamefont {F.}~\bibnamefont
  {Becattini}}, \bibinfo {author} {\bibfnamefont {J.}~\bibnamefont {Cleymans}},
  \bibinfo {author} {\bibfnamefont {A.}~\bibnamefont {Keranen}}, \bibinfo
  {author} {\bibfnamefont {E.}~\bibnamefont {Suhonen}}, \ and\ \bibinfo
  {author} {\bibfnamefont {K.}~\bibnamefont {Redlich}},\ }\href {\doibase
  10.1103/PhysRevC.64.024901} {\bibfield  {journal} {\bibinfo  {journal} {Phys.
  Rev. C}\ }\textbf {\bibinfo {volume} {64}},\ \bibinfo {pages} {024901}
  (\bibinfo {year} {2001})},\ \Eprint {http://arxiv.org/abs/hep-ph/0002267}
  {arXiv:hep-ph/0002267} \BibitemShut {NoStop}%
\bibitem [{\citenamefont {Becattini}\ \emph {et~al.}(2014)\citenamefont
  {Becattini}, \citenamefont {Grossi}, \citenamefont {Bleicher}, \citenamefont
  {Steinheimer},\ and\ \citenamefont {Stock}}]{Becattini:2014hla}%
  \BibitemOpen
  \bibfield  {author} {\bibinfo {author} {\bibfnamefont {F.}~\bibnamefont
  {Becattini}}, \bibinfo {author} {\bibfnamefont {E.}~\bibnamefont {Grossi}},
  \bibinfo {author} {\bibfnamefont {M.}~\bibnamefont {Bleicher}}, \bibinfo
  {author} {\bibfnamefont {J.}~\bibnamefont {Steinheimer}}, \ and\ \bibinfo
  {author} {\bibfnamefont {R.}~\bibnamefont {Stock}},\ }\href {\doibase
  10.1103/PhysRevC.90.054907} {\bibfield  {journal} {\bibinfo  {journal} {Phys.
  Rev. C}\ }\textbf {\bibinfo {volume} {90}},\ \bibinfo {pages} {054907}
  (\bibinfo {year} {2014})},\ \Eprint {http://arxiv.org/abs/1405.0710}
  {arXiv:1405.0710 [nucl-th]} \BibitemShut {NoStop}%
\end{thebibliography}%

\end{document}